\documentclass[preprint,12pt]{elsarticle}

 \usepackage{graphics}
 \usepackage{graphicx}
\usepackage{epsfig}

\usepackage{amssymb}

\usepackage{amssymb}

\journal{Experimental Thermal and Fluid Science}

\begin{document}

\begin{frontmatter}



\title{Theoretical Study of Steam Condensation Induced Water Hammer Phenomena in Horizontal Pipelines}

\


\author{ Ime Ferenc Barna$^{*,1}$ and Attila Rik\'ard Imre $^{2}$}

\address{1) Wigner Research Centre of the Hungarian Academy of Sciences, \\
1121 Budapest, Konkoly Thege \'ut 29-33, Hungary \\
Tel: +36-1-392-2222/3504, Fax: +36-1-395-9151  \\ 
2) Energie Research Center of the Hungarian Academy of Sciences, \\
Konkoly Thege ut 29-33, 1121, Budapest, Hungary}

\begin{abstract}

We investigate steam condensation induced water hammer (CIWH) phenomena and present new theoretical results. We use the WAHA3 model based on two-phase flow six first-order partial differential equations that present one dimensional, surface averaged mass, momentum and energy balances. A second order accurate high-resolution shock-capturing numerical scheme was applied with different kind of limiters in the numerical calculations. The applied two-fluid model shows some similarities to Relap5 which is widely used in the nuclear industry to simulate nuclear power plant accidents. This model was validated  with different CIWH experiments which were performed in the PMK-2 facility, which is a full-pressure thermo-hydraulic  model of the nuclear power plant of VVER-440/312 type in the Energy Research Center of the Hungarian Academy of  Sciences in Budapest and in the Rosa facility in Japan. In our recent study we show the first part of a planned large database which will give us the upper and lower flooding mass flow rates for various pipe geometries where CIWH can happen. Such a reliable database would be a great help for future reactor constructions and scheming. 
\end{abstract}


\end{frontmatter}



\section{Introduction }

\let\thefootnote\relax\footnote{*barna.imre@wigner.mta.hu}
\label{}

Safety of nuclear reactors is a fundamental issue.
Nuclear and thermo-hydraulic processes in the active zone
of modern reactors are well known and well-controlled,
explosions are out of question. However, violent unwanted
thermo-hydraulic transients in the primer loop may cause
serious deformation or pipe breakage. Such an unplanned
transient is the CIWM. In thermal loops of atomic reactors
or in other pipelines where water steam and cold water can
mix, quick and dangerous transients can happen causing
pressure surges which mean high financial expenses or
even cost human lives.
In the following we will present the WAHA3 model \cite{tis1},
 which is a complex physical model suitable to
simulate various quick transients in single and two-phase
flows, such as ideal gas Riemann problem, critical flow of
ideal gas in convergent-divergent nozzle, rapid
depressurization of hot liquid from horizontal pipes and
column separation water hammer or even CIWH.
In the last two decades the nuclear industry
developed a few complex two-phase flow-codes like
RELAP5  \cite{kars}, TRAC  \cite{trac} or CATHARE \cite{cath} which are feasible to solve safety analysis
of nuclear reactors and model complicated two-phase flow
transients.
The model, WAHA3 has some similarities with
RELAP5. This means that the conservation equations are the
same but the applied correlations are partially different.
The main difference between the above mentioned models
and our WAHA3 code is basically the applied numerical
scheme; other commercial codes have a ratio of spatial and
time resolution $\Delta x /\Delta t  $ which describes usual flow
velocities. WAHA3, however is capable of capturing shock
waves and describe pressure waves which may propagate
quicker than the local speed of sound. As a second point
WAHA3 has a quick condensation model which is not
available for RELAP5 and CATHARE.


\section{Experimental Facility}
This is a theoretical and numerical study, the used
experimental facilities PMK2, and ROSA were mentioned
in our former studies where validations were also made\cite{barna1,barna2}.


\section{Numerical Scheme}
There are large number of different two-phase flow models
with different levels of complexity
\cite{stew,men,ishibi}  which are all based on gas dynamics and
shock-wave theory. In the following we present the one
dimensional six-equation equal-pressure two-fluid model.
The density, momentum and energy balance equations for
both phases are the following:

\begin{equation}
\frac{\partial A (1-\alpha) \rho_l  }{\partial t } + \frac{\partial A (1-\alpha) \rho_l (v_l - w) }{\partial x } = -A\Gamma_g   
\end{equation}
\begin{equation}
\frac{\partial A \alpha \rho_g  }{\partial t } + \frac{\partial A \alpha \rho_g (v_g - w) }{\partial x } = A\Gamma_g   
\end{equation} 
\begin{eqnarray}
\frac{\partial A(1- \alpha) \rho_l v_l  }{\partial t } +  \frac{\partial A (1-\alpha) \rho_l v_l(v_l - w) }{\partial x } +    
A(1-\alpha)\frac{\partial p}{\partial x } - A \cdot CVM    \nonumber  \\  - Ap_i\frac{\partial \alpha}{\partial x } = AC_i |v_r| v_r - A\Gamma_g v_l 
 + A(1-\alpha)\rho_l cos(\theta) - A F_{l,wall}  
\end{eqnarray}
\begin{eqnarray}
\frac{\partial A\alpha \rho_g v_g  }{\partial t } +  \frac{\partial A \alpha \rho_g v_g(v_g - w) }{\partial x } +    
A\alpha\frac{\partial p}{\partial x } - A \cdot CVM  - Ap_i\frac{\partial \alpha}{\partial x } =  \nonumber \\  AC_i |v_r| v_r -   A\Gamma_g v_g 
 + A\alpha\rho_g cos(\theta) - A F_{g,wall}  
\end{eqnarray}
\begin{eqnarray}
\frac{\partial A(1- \alpha) \rho_l e_l  }{\partial t } +  \frac{\partial A (1-\alpha) \rho_l e_l(v_l - w) }{\partial x } +    
p \frac{\partial A (1-\alpha)}{\partial t}  + \nonumber \\  \frac{\partial A (1-\alpha) p(v_l - w) }{\partial x }
 = A Q_{il}    -A \Gamma_g (h_l + v_l^2/2)  + A (1-\alpha )\rho_l v_l g cos(\theta) 
\end{eqnarray}
\begin{eqnarray}
\frac{\partial A\alpha \rho_g e_g  }{\partial t } +  \frac{\partial A \alpha \rho_g e_g(v_g - w) }{\partial x } +    
p \frac{\partial A \alpha}{\partial t}  + \nonumber \\  \frac{\partial A \alpha p(v_g - w) }{\partial x }
 = A Q_{ig}    -A \Gamma_g (h_g + v_g^2/2)  + A \alpha \rho_g v_g g cos(\theta) 
\end{eqnarray}

Index l refers to the liquid phase and the index g to the gas phase. 
Nomenclature and variables are described at the end of the manuscript. Left hand side of the
equations contains the terms with temporal and spatial derivatives.   
 Hyperbolicity of the equations is ensured with the virtual mass term CVM and with the 
interfacial term (terms with $p_i$). Terms on the right hand side are terms describing 
the inter-phase heat, mass  (terms with $\Gamma_g$)
vapor generation rate, volumetric heat fluxes $Q_{ij}$ ,
momentum transfer (terms with
$C i$ ), wall friction
$F_{g , wall}$ , and gravity terms. Modeling of the inter-phase
heat, mass and momentum exchange in two-phase models
relies on correlations which are usually flow-regime
dependent.
The system code RELAP5 has a very sophisticated flow
regime map with a high level of complexity. WAHA3
however has the most simple flow map with dispersed and
horizontally stratified regimes only. The uncertainties of
steady-state correlations in fast transients are very high.
A detailed analysis of the source terms can be found
elsewhere \cite{tis2}.
Two additional equation of states(eos) are needed to
close the system of Eqs. (1-6.) Here the subscript k can
have two values ‘l’ for liquid phase, and ‘g’ for gas phase
\begin{equation}
\rho_k = \left( \frac{\partial \rho_k}{\partial p} \right)_{u_k} dp + 
\left( \frac{\partial \rho_k}{\partial u_k} \right)_p du_k
\end{equation}

Partial derivatives in Eq. 7 are expressed using pressure
and specific internal energy as an input. The table of water
and steam properties was calculated with a software from
UCL \cite{sey}.
The system of Eqs. (1-6) represents the conservation laws
and can be formulated in the following vectorial form
\begin{equation}
\underline{\underline{A}} \frac{\partial \overline{ \Psi}}{\partial t} + \underline{\underline{A}} \frac{\partial\overline{ \Psi}}{\partial x} = \overline{S} 
\end{equation}
where $\overline{ \Psi}$  represents a vector of the non-conservative variables 
 
$\overline{ \Psi}(p,\alpha,v_l,v_g,u_l,u_g)$ and $ \underline{\underline{A}},\underline{\underline{B}} $ are 6-times-6 matrices and 
$\overline{S} $
is the source vector of
non-differential terms. These three terms can be obtained
from Eq. (1-6) with some algebraic manipulation.
In this case the system eigenvalues which represent
wave propagation velocities are given by the determinant $
det( \underline{\underline{A}} - \lambda \underline{\underline{B}})
$. 
An improved characteristic upwind discretization method is used to solve   the hyperbolic
equation system (Eq. 8). The problem is solved with the
combination of the first- and second-order accurate
discretization scheme by the so-called flux limiters to
avoid numerical dissipation and unwanted oscillations
which appear in the vicinity of the non-smooth solutions.
Exhaustive details about the numerical scheme can be
found in the work of \cite{lev}.


\section{Results and Discussion} 
In our recent study we investigated pipe lines with three
different diameters (D= 10,20,50 cm) with three different
tube aspect ratio (L/D = 25,50,75) and with three different
pressures (p=10, 20,40 bar).
These are physically relevant geometries with pressures
values which are interesting in various nuclear facilities.
Table I presents these system parameters with the minimal
and the maximal mass flow rates in between CIWH events
happen.
For a better transparency these results are presented on Fig
1,2 and 3 for the different pipe diameters. With this useful
representation we can immediately see the dangerous
CIWH range between the upper and lower flooding mass
flow rates.
For completeness we explain additional technical
details of our investigations. In all calculations we used the
same nodalization in the sense that the actual length of the
node is equal to the actual pipe diameter. In all calculations
the same Courant-Friedrich-Levy limit was applied with
0.8. As numerical scheme the MINMOD limiter was used.
There are only two exceptions at D = 50, L/D = 50, 75 p
=20 bar.
The temperature of the cold water was fixed to 293 K.
Each presented system (e.g. D = 10 cm, L/D=25, p =20 bar
minimal mass flow rate) means at least 10 independent
calculations with slightly different mass flow parameters.
For the maximal flow a calculation takes 20 minutes or
even less but for the minimal flow rate one calculation
might take 20 hours. To determine if a CIWH event
happened we simple checked the pressure-time history
closed to the cold water inlet visually. If a sharp peak with
a 2 milisecond FWHM can be seen than we are in the
dangerous water hammer region. It is worth to note that,
our experience shows that there is a very sharp border at
both sides of the CIWH regime in this WAHA3 model.
Unfortunately, the curves in Figure 1,2,3 are not parallel
and cross each other which is a confusing problem at this
moment. As explanation we think to say that, with
additional very time consuming tuning of all the technical
parameters (limiter, CFL condition, nodalisation) some of
the border points could be slightly modified, but this was
not possible till now.


\begin{figure} 
\begin{center}
\scalebox{0.4}{\rotatebox{-90}{\includegraphics{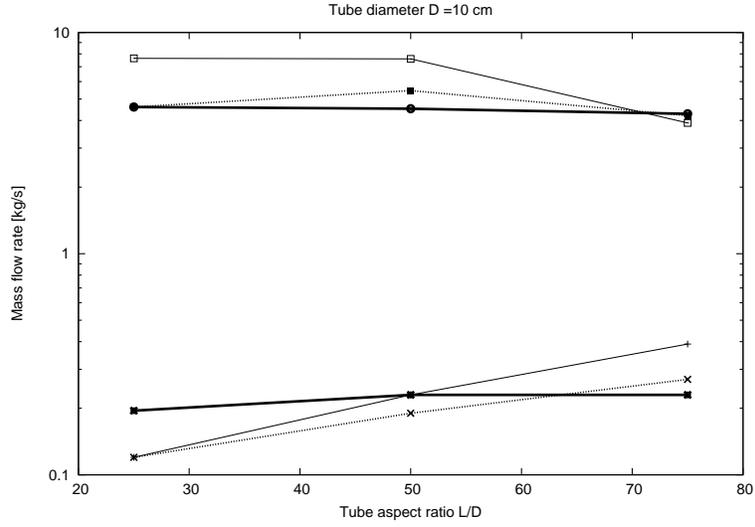}}} 
\caption{ The minimal and maximal mass flow rates for
the D=10 cm diameter pipelines for L/D = 25, 50, 75
tube aspect ratios.  The thin solid curve is for p = 10 bar,  the dashed 
is for = 20 bar, and the thick solid one is for p = 40 bar.
 } \label{otos}       
\end{center}
\end{figure}

\begin{figure} 
\begin{center}
\scalebox{0.4}{\rotatebox{-90}{\includegraphics{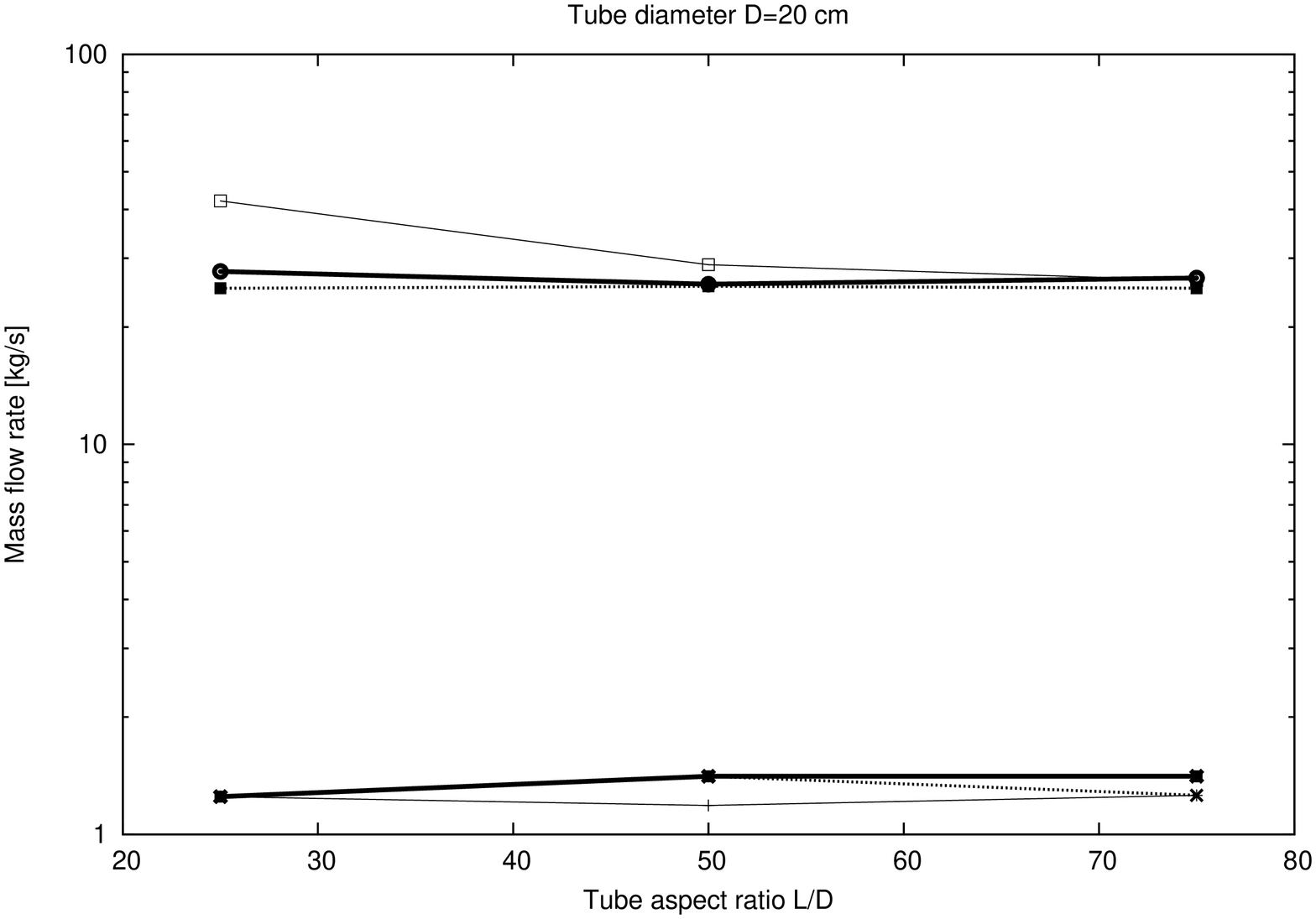}}} 
\caption{The minimal and maximal mass flow rates for
the D=20 cm diameter pipelines for L/D = 25, 50, 75
tube aspect ratios . The thin solid curve is for p = 10 bar,  the dashed 
is for = 20 bar, and the thick solid one is for p = 40 bar
 } \label{ootos}       
\end{center}
\end{figure}

\begin{figure} 
\begin{center}
\scalebox{0.4}{\rotatebox{-90}{\includegraphics{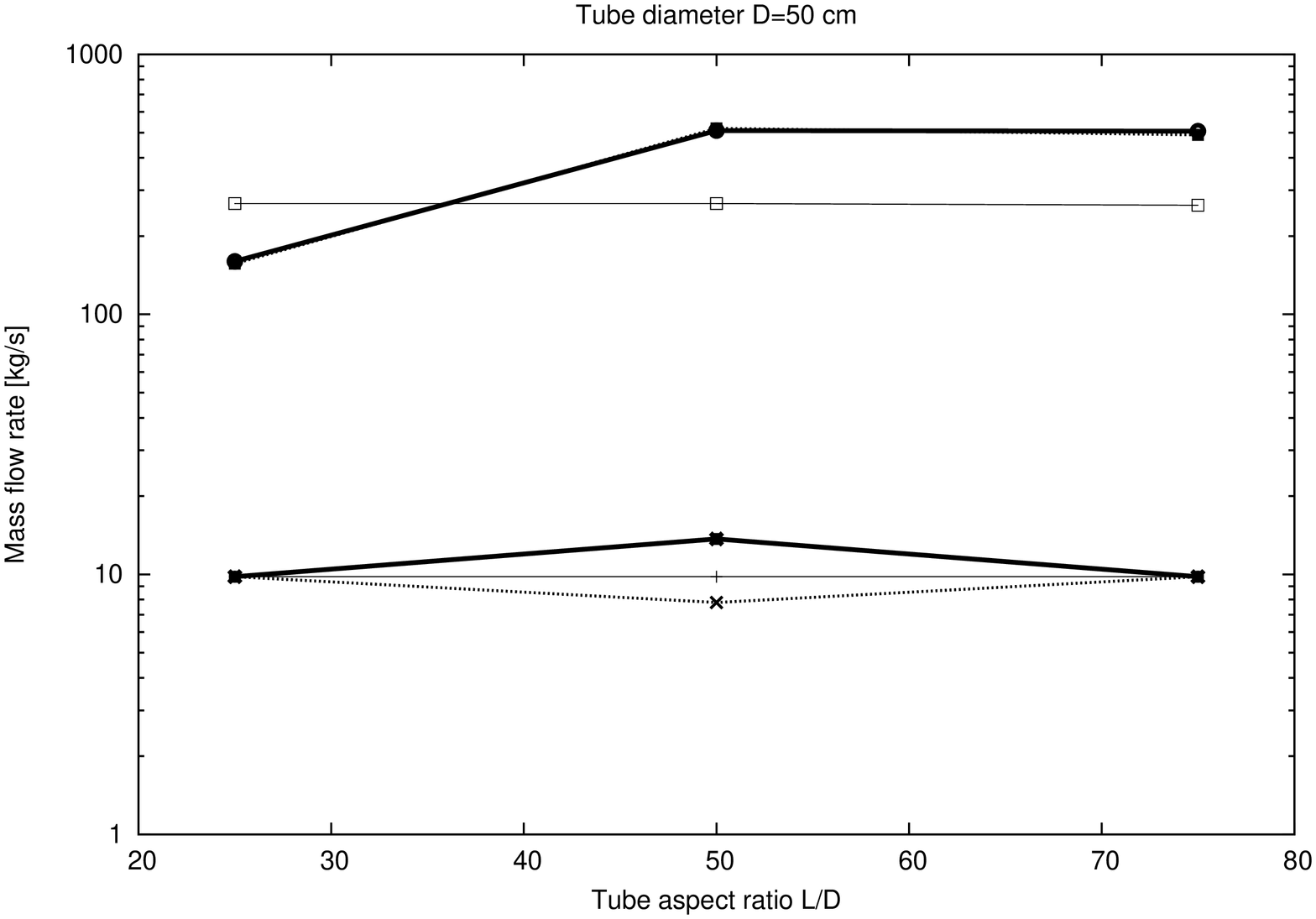}}} 
\caption{The minimal and maximal mass flow rates for
the D=50 cm diameter pipelines for L/D = 25, 50, 75
tube aspect ratios . Thin solid curve is for p = 10 bar, the dashed curve
is for = 20 bar, and the thick solid one is for p = 40 bar
 } \label{ootos}       
\end{center}
\end{figure}


\section{Conclusions}
We presented the WAHA3 numerical model which is
capable to describe supersonic two-phase flow transients in
pipe lines. After our former  CIWH studies \cite{barna1,barna2},
we presented now a database where the minimal and
maximal mass flow rates can be determined for large
number of flow systems. We plan to go further and enhance
our database for even larger number of systems.

\section{Acknowledgements}
We thank Prof. Dr. Iztok Tiselj (Jozef Stefan Institute,
Ljubljana, Slovenia) for his fruitful discussions and
valuable comments.



\newpage 

\begin{center}
\begin{tabular}{ |l|c|c| }
\hline
System   & Minimal mass  & Maximal mass    \\  
Parameters  &   flow rate  (kg/s) &  flow rate (kg/s)  \\ \hline \hline 
{\bf{D = 10 cm}} & &   \\ 
{\it{L/D = 25}} & &  \\ \hline 
p = 10 bar & 0.12& 7.64 \\  
p = 20 bar & 0.12& 4.60 \\   
p = 40 bar & 0.195& 4.60 \\  \hline    
{\it{L/D = 50}} & &  \\ \hline 
p = 10 bar & 0.23& 7.64 \\  
p = 20 bar & 0.19& 5.46 \\   
p = 40 bar & 0.23& 4.52 \\  \hline     
{\it{L/D = 75}} & &  \\ \hline 
p = 10 bar & 0.39& 3.90 \\  
p = 20 bar & 0.27& 4.21 \\   
p = 40 bar & 0.23& 4.29 \\  \hline    \hline 
{\bf{D = 20 cm}} & &   \\ 
{\it{L/D = 25}} & &  \\ \hline 
p = 10 bar & 1.25& 42.08 \\  
p = 20 bar & 1.25& 25.12 \\  
p = 40 bar & 1.25& 27.75 \\  \hline    
{\it{L/D = 50}} & &  \\ \hline 
p = 10 bar & 1.18& 28. 89 \\  
p = 20 bar & 1.41& 25.43 \\   
p = 40 bar & 1.41& 25.75 \\  \hline     
{\it{L/D = 75}} & &  \\ \hline 
p = 10 bar & 1.26& 26.38 \\  
p = 20 bar & 1.26& 25.12 \\   
p = 40 bar & 1.41& 26.70 \\  \hline   \hline 
{\bf{D = 50 cm}} & &   \\ 
\vspace*{-2mm}{\it{L/D = 25}} & &  \\ \hline 
p = 10 bar & 9.8& 266. 5 \\  
p = 20 bar & 9.8& 156.8 \\   
p = 40 bar & 9.8& 160.0 \\  \hline    
{\it{L/D = 50}} & &  \\ \hline 
p = 10 bar & 9.8& 266.55 \\  
p = 20 bar & 7.8& 519.4\\   
p = 40 bar & 13.7& 509.0\\  \hline    
{\it{L/D = 75}} & &  \\ \hline 
p = 10 bar & 9.8& 262.66 \\  
p = 20 bar & 9.8& 490.0 \\  
p = 40 bar & 9.8& 505.6\\  \hline    

\end{tabular} \\   
\vspace*{5mm }
Table I. The minimal and maximal mass flow rates for the investigated systems. 
\end{center}

 \section{Nomenclature} 
\begin{tabular}{ l l }
A    &         pipe cross section ($m^2$ ) \\
$C_i$  &  		internal friction coefficient	($kg/m$) \\
CVM     &   virtual mass term $(N/ m^3)$ \\
$e_i $  & specific total energy $[e = u + v^2 /2] (J/kg)$ \\
$F_{g,wall}$ &  wall friction per unit volume $(N/m)$ \\ 
 g & gravitational acceleration $(m/s^2)$ \\ 
$h_i$  & specific enthalpy    $[h = u + p/ \rho]$	(J/kg) \\ 
p & Pressure $(Pa)$ \\
$p_i$ &  interfacial pressure $p_i   = p\alpha (1- \alpha) $  (Pa)   \\ 
$Q_{ij} $&  interf.-liq./gas heat transf. per vol. rate $(W/m^3 )$  \\ 
t  & 	time $(s)$ \\ 
$u_i$ & 	specific internal energy $(J/kg)$ \\
$v_i$ & velocity $(m/s)$ \\ 
$v_r$ & relative velocity $(v_r = v_g - v_f) (m/s)$ \\ 
w & pipe velocity in flow direction $(m/s)$ \\ 
x & spatial coordinate $(m)$ \\ 
Greek letters &  \\ 
$\alpha$  & vapour void fraction \\ 
$\Gamma_g$  & vapour generation rate $(kg/m^3)$ \\  
$\rho_i$ & density $(kg/m^3)$ \\ 
$\theta$  &  pipe inclination $(degree)$ \\ 
Subscripts & \\ 
l & liquid phase  	\\ 
g & gas phase  
\end{tabular}
\end{document}